\documentclass[english,french,british,aip,pof]{revtex4-1}
\usepackage[T1]{fontenc}
\usepackage[latin9]{inputenc}
\usepackage{xcolor}
\usepackage{pdfcolmk}
\usepackage{bm}
\usepackage{amsmath}
\usepackage{amssymb}
\usepackage{graphicx}
\PassOptionsToPackage{normalem}{ulem}
\usepackage{ulem}
\makeatletter

\providecolor{lyxadded}{rgb}{0,0,1}
\providecolor{lyxdeleted}{rgb}{1,0,0}

 \@ifundefined{textcolor}{}
 {%
   \definecolor{BLACK}{gray}{0}
   \definecolor{WHITE}{gray}{1}
   \definecolor{RED}{rgb}{1,0,0}
   \definecolor{GREEN}{rgb}{0,1,0}
   \definecolor{BLUE}{rgb}{0,0,1}
   \definecolor{CYAN}{cmyk}{1,0,0,0}
   \definecolor{MAGENTA}{cmyk}{0,1,0,0}
   \definecolor{YELLOW}{cmyk}{0,0,1,0}
 }

\@ifundefined{showcaptionsetup}{}{%
 \PassOptionsToPackage{caption=false}{subfig}}
\usepackage{subfig}
\makeatother

\usepackage{babel}
\addto\extrasfrench{%
   \providecommand{\fg}{\ifdim\lastskip>\z@\unskip\fi~\frqq}%
}

\begin{document}

\title{Analysis of transient growth using an orthogonal decomposition of
the velocity field in the Orr-Sommerfeld Squire equations}

\author{Marc Buffat}

\email{marc.buffat@univ-lyon1.fr}

\homepage{www.ufrmeca.univ-lyon1.fr/~buffat}

\author{Lionel Le Penven}

\affiliation{Université de Lyon,\\
 Université Claude-Bernard Lyon 1/ CNRS/ Ecole Centrale de Lyon/ Insa
de Lyon}

\address{Laboratoire de Mécanique des Fluides et Acoustique LMFA, 43, bd du
11 nov. 1918, 69622 Villeurbanne, FRANCE}

\date{Internal report LMFA, July 2012}
\begin{abstract}
Despite remarkable accomplishment, the classical hydrodynamic stability
theory fails to predict transition in wall-bounded shear flow. The
shortcoming of this modal approach was found 20 years ago and is linked
to the non-orthogonality of the eigenmodes of the linearised problem,
defined by the Orr Sommerfeld and Squire equations. The associated
eigenmodes of this linearised problem are the normal velocity and
the normal vorticity eigenmodes, which are not dimensionally homogeneous
quantities. Thus non-orthogonality condition between these two families
of eigenmodes have not been clearly demonstrated yet. Using an orthogonal
decomposition of solenoidal velocity fields, a velocity perturbation
is expressed as an $L_{2}$ orthogonal sum of an Orr Sommerfeld velocity
field (function of the perturbation normal velocity) and a Squire
velocity field (function of the perturbation normal vorticity). Using
this decomposition, a variational formulation of the linearised problem
is written, that is equivalent to the Orr Sommerfeld and Squire equations,
but whose eigenmodes consist of two families of velocity eigenmodes
(thus dimensionally homogeneous). We demonstrate that these two sets
are non-orthogonal and linear combination between them can produce
large transient growth. Using this new formulation, the link between
optimal mode and continuous mode transition will also be clarified,
as the role of direct resonance. Numerical solutions are presented
to illustrate the analysis in the case of thin boundary layers developing
between two parallel walls at large Reynolds number. Characterisations
of the destabilizing perturbations will be given in that case.
\end{abstract}


\maketitle

\section{Introduction}

Since the pioneer work of Reynolds (1883), the study of transition
in wall-bounded shear flow is still today the subject of intense research
(see e.g. Schmid\cite{SchmidAnnRev07} for a review). Despite remarkable
accomplishment, the classical hydrodynamic stability theory fails
to predict transition in wall-bounded shear flow. The shortcoming
of this modal approach was found 20 years ago (see e.g. Trefethen
et al\cite{Trefethen1993} and Schmid and Henningson\cite{Schmid2001})
and is linked to the non-orthogonality of the eigenmodes of the linearised
problem, defined by the Orr Sommerfeld and Squire (O-S) equations.
Onset of transition in wall-bounded flows is characterised by subcritical
instability, i.e. instability that occurs below the critical Reynolds
number predicted by O-S equations. A theory for subcritical instability
that has received a great deal of attention is the transient growth
(Butler and Farrell\cite{Butler1992}). It is based on the observation
that a general initial wave other than a pure eigenmode may undergo
transient growth, even though all eigenmodes are decaying. Among this
form of initial perturbations, the one which yields the largest amplification
is referred to as being ``optimal''. The transient growth theory
emphasises the linear nature of the non modal amplification mechanism
(Trefethen et al\cite{Trefethen1993} and Henningson et al\cite{Henningson1994}).

The evolution of infinitesimal perturbations with two homogeneous
and one inhomogeneous directions is described by the linearised incompressible
Navier-Stokes equations. By using a velocity-vorticity formulation,
the linearised equations reduce to the classical Orr-Sommerfeld and
Squire equations. The associated matrix operators are non-normal and
large transient growth of energy is possible, even if all eigenvalues
are confined to the stable half-plane. Using the transient growth
theory, the initial conditions that will reach the maximum possible
amplification at a given time can be determined (Butler and Farrell\cite{Butler1992}
and Anderson et al\cite{AndersonPhysF99}) and are called optimal
disturbances\emph{.} For the Blasius boundary layer, the optimal perturbations
are stationary streamwise vortices inside the boundary layer, periodic
in the spanwise direction with wavelength of the order of 2 times
the boundary layer thickness (Luchini\cite{LuchiniJFM00}). The associated
optimal responses are large steady streamwise streaks, that are created
due to lift-up of mean momentum by the initial cross stream velocity
perturbations. These optimal streaks have been observed experimentally
by perturbing the boundary layer with the means of spanwise periodic
array of small cylindrical roughness elements fixed on the wall (Fransson
et al\cite{Fransson2004}).

Streamwise streaks are also observed in boundary layer in the presence
of free-stream turbulence. These streaks forced by free-stream turbulence
typically slowly oscillate in the boundary layer and are called Klebanoff
modes\emph{.} Destabilisation of these streaks can then induce a bypass
transition in boundary layer. Using the notion of continuous mode
transition, Durbin and Wu\cite{Durbin2007} attribute the creation
of these streaks to penetrating (or low frequency) modes in the continuous
spectrum of the O-S equations. These penetrating modes generate perturbation
jets (streaks) by lift-up of fluid into the upper portion of the layer.
Then non penetrating modes (with higher frequency) trigger breakdown,
ultimately producing a turbulent transition. To quantify the forcing
by free-stream turbulence, Zaki and Durbin\cite{Zaki2005} define
a coupling coefficient, related to the forcing term in the Squire
equation, that characterize modal penetration of free-stream turbulence
inside the boundary layer. They also attribute the growth of disturbances
to exact resonance between Orr-Sommerfeld and Squire eigenmodes.

In this paper we will revisit the classical O-S equations using an
orthogonal decomposition of solenoidal velocity fields, based on the
normal velocity and vorticity component. This allows to analyse the
linearised Navier-Stokes equations (equivalent to the O-S equations)
in term of two dimensionally homogeneous families of velocity eigenmodes
instead of the classical (not dimensionally homogeneous) eigenmodes
for the normal velocity and vorticity (Schmid and Henningson\cite{Schmid2001}).
Using this formulation, the link between optimal mode and continuous
mode transition will be clarified. We will also address the question
of the role of direct resonance as it was suggested by the asymptotic
analysis of Hultgren et al\cite{Hultgren1981}, and emphasis in Zaki
and Durbin\cite{Zaki2005}.

This paper is organised as follows. Section \ref{sec:decomposition}
contains the description of the orthogonal decomposition used for
the analysis. In Section \ref{sec:Wall-bounded-shear}, this decomposition
is applied to the linear stability analysis of parallel flow in bounded
domain. Numerical solutions will be presented in Section \ref{sec:Numerical-solution}
to illustrated the analysis in the case of thin boundary layers developing
between two parallel walls at large Reynolds number. Discussion and
concluding remarks are then presented in Section \ref{sec:Discussion}.

\section{Orthogonal decomposition of solenoidal fields\label{sec:decomposition}}

Orthogonal decomposition of solenoidal fields, based on complete representations,
have proved invaluable in studies of hydrodynamic stability (Chandrasekhar\cite{chandrasekha81}),
fluid turbulence (Holmes et al\cite{lumley96}) and magneto-hydrodynamic
turbulence (Turner\cite{TurnerJPhysA07}). For instance, representation
of solenoidal vector field by poloidal and toroidal potentials (Warner\cite{warner72})
is a useful analytic technique in many problems of mechanics and electromagnetics.
Likewise, in stratified homogeneous turbulence, the formalism of Craya
can be used to decompose the velocity field in the Fourier space into
a vortical component and an orthogonal wavy component (Herring\cite{herring74}).

Le Penven and Buffat\cite{lepenven2012} proposed a general orthogonal
decomposition for solenoidal vector fields expressed in terms of projections
of the velocity and vorticity fields on an arbitrary direction in
space. For doubly-periodic flows with one direction of inhomogeneity
$\mathbf{e}_{y}$ (called the normal direction), Buffat et al\cite{Buffat2009}
derived an explicit form of this decomposition using the Helmholtz-Hodge
theorem (Chorin and Marsden\cite{chorin00}). The solenoidal velocity
field is expanded in Fourier series in the streamwise and spanwise
directions in the form:

\begin{equation}
\mathbf{u}(x,y,z,t)=\sum_{m=-\infty}^{\infty}\sum_{p=-\infty}^{\infty}\mathbf{u}^{mp}(y,t)\, e^{\bm{\imath}(\alpha x+\beta z)}\label{eq3}
\end{equation}
where $\mathbf{u}^{mp}$ is the vector function of Fourier coefficients
associated with wave-numbers $\alpha$ and $\beta$. Each Fourier
mode $\mathbf{u}^{mp}$ can thus be decomposed as $\mathbf{u}^{mp}=\mathbf{u}_{os}^{mp}+\mathbf{u}_{sq}^{mp}$,
where $\mathbf{u}_{os}^{mp}$ and $\mathbf{u}_{sq}^{mp}$ are respectively
functions of $v(y)$ and $\omega(y)$ that are the Fourier modes of
the normal velocity and normal vorticity:

\begin{equation}
\mathbf{u}_{os}^{mp}=\left(\bm{\imath}\,\frac{\alpha}{k^{2}}\partial_{y}v\,,\, v\,,\,\bm{\imath}\,\frac{\beta}{k^{2}}\partial_{y}v\right)^{t}\,,\,\mathbf{u}_{sq}^{mp}=\left(-\bm{\imath}\,\frac{\beta}{k^{2}}\omega\,,\,0\,,\,\bm{\imath}\,\frac{\alpha}{k^{2}}\omega\right)^{t}\mbox{ with }k^{2}=\alpha^{2}+\beta^{2}\label{eq4}
\end{equation}
Both velocity and vorticity components verify orthogonality conditions:
$\mathbf{u}_{os}^{mp}.\mathbf{u}_{sq}^{mp}=0$ and $\left(\mathcal{D}_{mp}\times\mathbf{u}_{os}^{mp}\right).\left(\mathcal{D}_{mp}\times\mathbf{u}_{sq}^{mp}\right)=0$,
where $\mathcal{D}_{mp}$ is the gradient operator in Fourier space:
$\mathcal{D}_{mp}=\left(\begin{array}{ccc}
{\bf \bm{\imath}}\alpha & ,\partial_{y} & ,\bm{\imath}\beta\end{array}\right)^{t}$. An $L_{2}$ orthogonal decomposition for the velocity field, function
of the normal velocity and vorticity fields, can easily be derived
from this orthogonal decomposition of the Fourier components :

\begin{equation}
\mathbf{u}=\mathbf{u}_{sq}(v)+\mathbf{u}_{os}(\omega)\mbox{ with }<\mathbf{u}_{sq},\mathbf{u}_{os}>_{L_{2}}=0\mbox{ and }\nabla.\mathbf{u}_{sq}=\nabla.\mathbf{u}_{os}=0\label{eq1}
\end{equation}

This decomposition induces an $L_{2}$ orthogonal decomposition of
the solenoidal fields space $W$ into two subspaces $W^{os}$ and
$W^{sq}$, and a similar decomposition of the space $W_{mp}$ of divergence
free Fourier modes $\mathbf{u}^{mp}$ (i.e. with $\mathcal{D}_{mp}.\mathbf{u}_{mp}=0$).

In the following, the decomposition \eqref{eq1} will be applied to
parallel flows in a bounded domain. Linear stability theory of parallel
flows is generally formulated in terms of two scalar differential
equations: the Orr-Sommerfeld equation for the $y$ component of the
velocity $v$ and the Squire equation for the $y$ component of the
vorticity $\omega$ (Schmid and Henningson\cite{Schmid2001}). As
the two velocity fields in the decomposition \eqref{eq1} are defined
respectively by the normal component of velocity and the normal component
of vorticity, these two fields $\mathbf{u}_{os}$ and $\mathbf{u}_{sq}$
have been denominated as the Orr-Sommerfeld velocity (OS velocity)
and Squire velocity (SQ velocity).

\section{Shear flow in bounded domain\label{sec:Wall-bounded-shear}}

\begin{figure}
\begin{centering}
\resizebox{0.4\textwidth}{!}{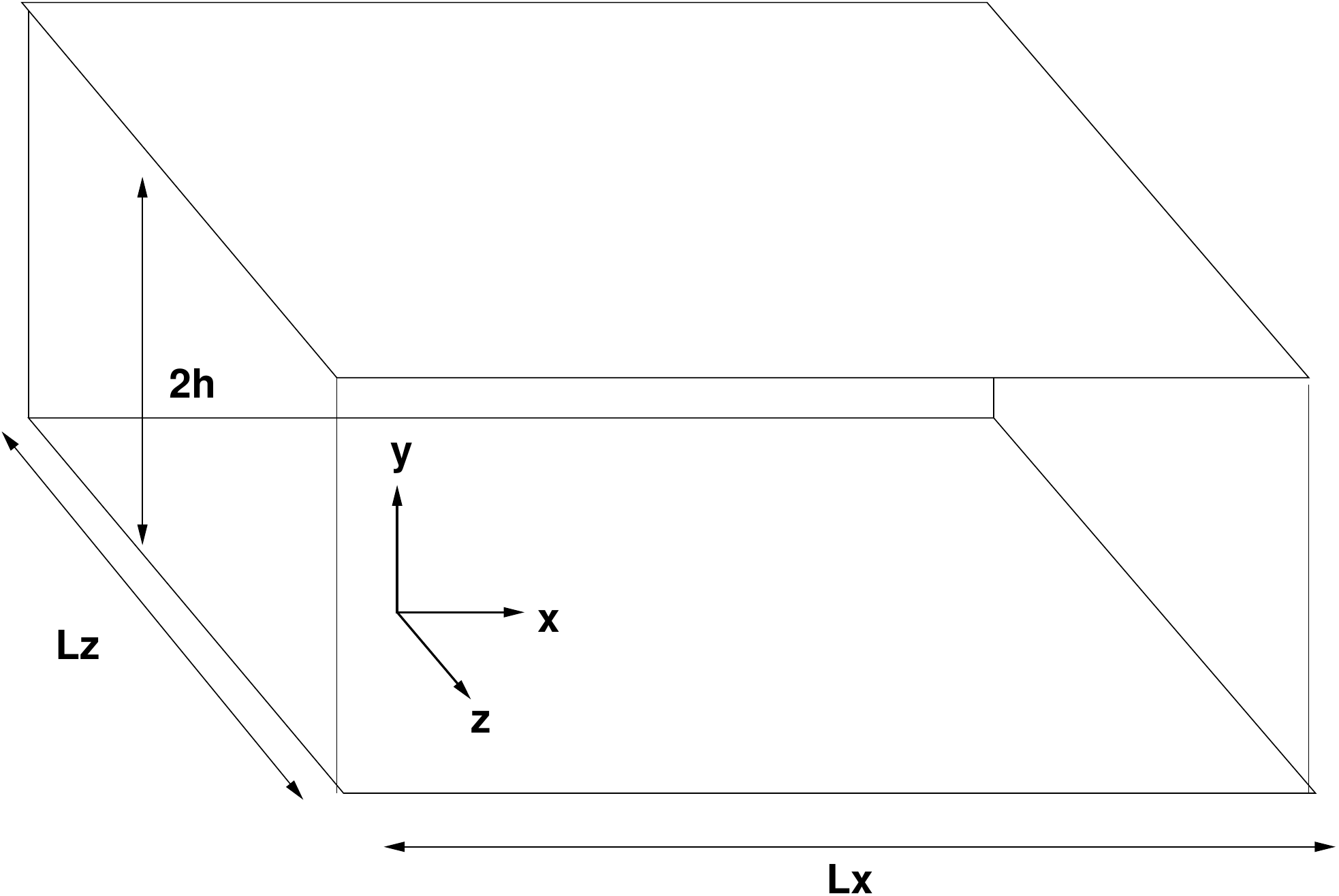}
\par\end{centering}

\caption{bounded domain$\Omega$\label{fig1}}
\end{figure}

A shear flow is considered in a parallelepiped $\Omega$ limited by
two parallel planes distant from $L_{y}=2h$ apart (see Figure\ref{fig1}).
As usual for local stability analysis, the base flow is assumed to
be parallel in the x-direction $\mathbf{U}_{b}=U_{b}(y)\,\mathbf{e}_{x}$
and the perturbation is expanded in Fourier series \eqref{eq3} with
homogeneous boundary conditions at $y=\pm h$.

Following Pasquarelli et al\cite{pasquarelli87}, a weak formulation
of the Navier-Stokes equations for the perturbation can be written
in divergence-free function space. By using an expansion in Fourier
series, the weak formulation for $\mathbf{u}_{mp}$ reads by virtue
of orthogonality of the trigonometric functions with respect to the
$L_{2}$ inner product $<.,.>_{L_{2}}$:

\begin{eqnarray}
\mbox{Find }\mathbf{u}_{mp}\in W_{mp}\mbox{ such that }\forall\mathbf{v}_{mp}\in W_{mp}\nonumber \\
<\partial_{t}\mathbf{u}_{mp}+\mathbf{U}^{b}.\mathcal{D}_{mp}\mathbf{u}_{mp}+\mathbf{u}_{mp}.\bm{\nabla}\mathbf{U}^{b}-Re^{-1}\mathcal{D}_{mp}^{2}\mathbf{u}_{mp},\mathbf{v}_{mp}>_{L_{2}} & = & <NL(\mathbf{u},\mathbf{u}),\mathbf{v}_{mp}>_{L_{2}}\label{eq7}
\end{eqnarray}
where $NL(\mathbf{u},\mathbf{u})$ represents the non linear terms.
We note that the pressure is eliminated from the formulation \eqref{eq7},
because the solution $\mathbf{u}_{mp}$ and its variations $\mathbf{v}_{mp}$
are solenoidal. Using the orthogonal decomposition \eqref{eq4}, the
weak formulation \eqref{eq7} is equivalent to:

\begin{eqnarray}
\mbox{Find }\mathbf{u}_{os}^{mp}(v)\in W_{mp}^{os}\mbox{ and }\mathbf{u}_{sq}^{mp}(\omega)\in W_{mp}^{sq}\mbox{ such that} &  & \forall\mathbf{v}_{os}(w),\,\forall\mathbf{v}_{sq}(\eta)\nonumber \\
<(\mathcal{M}+\mathcal{L})\mathbf{u}_{os}^{mp},\mathbf{v}_{os}>+\frac{\bm{\imath}\alpha}{k^{2}}<\partial_{y}\left(v\, d_{y}U^{b}\right),w> & = & <NL(\mathbf{u},\mathbf{u}),\mathbf{v}_{os}>\label{eq2a}\\
<(\mathcal{M}+\mathcal{L})\mathbf{u}_{sq}^{mp},\mathbf{v}_{sq}>+\frac{\bm{\imath}\beta}{k^{2}}<v\, d_{y}U^{b},\eta> & = & <NL(\mathbf{u},\mathbf{u}),\mathbf{v}_{sq}>\label{eq2b}
\end{eqnarray}
where $\mathcal{M}=\partial_{t}$ is the temporal operator and $\mathcal{L}=\bm{\imath}\alpha U^{b}\,-Re^{-1}\mathcal{D}_{mp}^{2}$
the diffusion and transport by the mean flow $U^{b}$ operator. The
linear parts of these equations are equivalent to the classical Orr-Sommerfeld/Squire
equations. As this weak formulation is equivalent to the virtual work
principle apply to the perturbation, the different terms of equations
(\ref{eq2a}-\ref{eq2b}) can also be interpreted in terms of transport,
diffusion and transfer of kinetic energy. By denoting $\mathcal{M}_{os}$
and $\mathcal{L}_{os}$ the OS projection of $\mathcal{M}$ and $\mathcal{L}$
(i.e. $\mathcal{M}_{os}\bm{u}=<\mathcal{M}\bm{u},\bm{v}_{os}>_{L_{2}}$),
and identically for $\mathcal{M}_{sq}$ and $\mathcal{L}_{sq}$, the
equivalent matrix form of equations (\ref{eq2a}-\ref{eq2b}) reads:

\begin{equation}
\underbrace{\left(\begin{array}{cc}
\mathcal{M}_{os}+\mathcal{L}_{os} & 0\\
0 & \mathcal{M}_{sq}+\mathcal{L}_{sq}
\end{array}\right)\left(\begin{array}{c}
\mathbf{u}_{os}^{mp}\\
\mathbf{u}_{sq}^{mp}
\end{array}\right)}_{transport\,\&\, diffusion}+\underbrace{\left(\begin{array}{c}
<\frac{\bm{\imath}\alpha}{k^{2}}\partial_{y}\left(v\, d_{y}U^{b}\right),w>\\
<\frac{\bm{\imath}\beta}{k^{2}}\left(vd_{y}U^{b}\right),\eta>
\end{array}\right)}_{transfer\, of\, energy\, from\, U^{b}}=\underbrace{\left(\begin{array}{c}
<NL(\mathbf{u},\mathbf{u}),\mathbf{v}_{os}>\\
<NL(\mathbf{u},\mathbf{u}),\mathbf{v}_{sq}>
\end{array}\right)}_{redistribution}\label{eq8}
\end{equation}
In these equations, the term that represents the transfer of energy
between the base flow $U^{b}$ and the perturbation depends only on
Orr-Sommerfeld velocity and consists in two parts: the first, proportional
to the longitudinal wavenumber $\alpha$, in the first (Orr-Sommerfeld)
equation and the second, proportional to the transverse wavenumber
$\beta$, in the second (Squire) equation. If $\alpha\gg\beta$, then
the Orr-Sommerfeld perturbation is essentially a two-dimensional perturbation.
Energy can be transferred from the base flow to the streamwise Orr-Sommerfeld
velocity component, which can then increase with time to develop Tolmien
Schlichting (TS) waves. In that case, large Squire velocity cannot
be created by a linear mechanism. If $\alpha\ll\beta$, then the Orr-Sommerfeld
perturbation is essentially a streamwise vortex. Energy can be transferred
from the base flow to streamwise Squire velocity to develop large
streamwise Squire velocity component. This mechanism is interpreted
as the lift-up effect of a fluid particle by the normal velocity (Schmid
and Henningson\cite{Schmid2001}). As the transfer term does not depend
on Squire velocity, in absence of Orr-Sommerfeld velocity $\mathbf{u}_{os}^{mp}$
or if the spanwise wave-number $\beta$ is small, the Squire velocity
$\mathbf{u}_{sq}^{mp}$ is always damped if the non-linear right-hand
side is neglected.

\subsection{Temporal stability analysis\label{sub:Temporal-stability-analysis}}

To solve the linearised equations \eqref{eq8}, wavelike solutions
are sought of the form\foreignlanguage{english}{ $\mathbf{u}(\mathbf{x},t)=\mathbf{u}^{mp}(y)\, e^{\bm{\imath}(\alpha x+\beta z-\lambda t)}$}
where $\lambda$ is a complex frequency. The problem reduces to an
eigenvalue problem: $\mathbf{L}\mathbf{\tilde{u}}=-\bm{\imath}\lambda\mathbf{\tilde{u}}$.
The associated linear operator $\mathbf{L}$ is nonnormal, and the
vector eigenfunctions\foreignlanguage{french}{ $\mathbf{\tilde{u}}^{l}(y)$}
associated with the eigenvalues $\lambda_{l}$ do not form an orthogonal
set. However, because the operator $\mathbf{L}$ is compact, in bounded
domains it has infinitely many isolated eigenvalues and the set $E_{mp}$
of normalized eigenvectors $\tilde{\mathbf{u}}^{l}(y)$ forms a complete
set in the solution space $W_{mp}$ (Prima and Habetler\cite{prima69}).
By using the orthogonal decomposition \eqref{eq1} on the eigenvectors
$\tilde{\mathbf{u}}^{l}=\mathbf{u}_{os}^{l}(v)+\mathbf{u}_{sq}^{l}(\omega)$,
the eigenvalue problem can be rewritten in matrix form:

\begin{equation}
\left[\begin{array}{cc}
\mathcal{L}_{os}+\mathcal{C}_{os} & 0\\
\mathcal{C}_{sq} & \mathcal{L}_{sq}
\end{array}\right]\left[\begin{array}{c}
\mathbf{u}_{os}^{l}\\
\mathbf{u}_{sq}^{l}
\end{array}\right]=-\bm{\imath}\lambda_{l}\left[\begin{array}{c}
\mathcal{M}_{os}\\
\mathcal{M}_{sq}
\end{array}\right]\left[\begin{array}{c}
\mathbf{u}_{os}^{l}\\
\mathbf{u}_{sq}^{l}
\end{array}\right]\label{eq9}
\end{equation}
which is equivalent to the classical Orr\textendash{}Sommerfeld/Squire
eigenvalue problem for the normal velocity $v$ and the normal vorticity
$\omega$. The main interest of the formulation \eqref{eq9} compared
to the classical O-S eigenvalue problem, is the dimensional homogeneity
of the two components $\mathbf{u}_{os}^{l}(v)$ and $\mathbf{u}_{sq}^{l}(\omega)$
of the eigenvectors. This allows one to define two distinct velocity
eigenvectors subsets $E_{mp}^{+}$ and $E_{mp}^{-}$ (with $E_{mp}^{+}\cup E_{mp}^{-}=E_{mp}$)
depending on their orthogonality with $W_{mp}^{os}$ (the space of
OS velocity Fourier modes):
\begin{enumerate}
\item $E_{mp}^{-}=\left\{ \mathbf{\tilde{u}}^{-,l}/\left\Vert \mathbf{\tilde{u}}^{-,l}\right\Vert =1\mbox{ and }\mathbf{u}_{os}^{-,l}=0\right\} $
includes velocity eigenmodes with no OS velocity components, i.e.
$E_{mp}^{-}\subset W_{mp}^{sq}\perp W_{mp}^{os}$,
\item $E_{mp}^{+}=\left\{ \tilde{\mathbf{u}}^{+,l}/\left\Vert \mathbf{\tilde{u}}^{+,l}\right\Vert =1\mbox{ and }\mathbf{u}_{os}^{+,l}\neq0\right\} $
includes velocity eigenmodes with OS velocity components, i.e. $E_{mp}^{+}\cap W_{mp}^{os}\neq\textrm{Ø}$.
\end{enumerate}
The subset $E_{mp}^{+}$ contains the eigenmodes of the Orr-Sommerfeld
eigenvalues problem and includes the classical longitudinal Tolmien
Schlichting waves. Meanwhile, the subset $E_{mp}^{-}$ contains the
eigenvalues of the homogeneous Squire equation. As the linear operator
$\mathcal{L}_{sq}$ associated with the set $E_{mp}^{-}$ is a transport-diffusion
operator, the velocity eigenmodes $\tilde{\mathbf{u}}^{-,l}$ are
always damped. Moreover $\mathcal{L}_{sq}$ is a compact linear operator
in $W_{mp}^{sq}$ and the set $E_{mp}^{-}$ forms a complete set in
$W_{mp}^{sq}$. Due to the linear coupling operator $\mathcal{C}_{sq}$
(with $\mathcal{C}_{sq}\mathbf{u}_{os}=<\bm{\imath}\frac{\beta}{k^{2}}v_{os}d_{y}U^{b},\eta>_{L_{2}}$),
which represents the transfer of energy from the base flow to the
SQ velocity by the mediation of an OS velocity, the two subsets $E_{mp}^{+}$
and $E_{mp}^{-}$ are not $L_{2}$-orthogonal, except for zero transverse
wave-number $(\beta=0)$. In that case, the set $E_{mp}^{+}$ is a
complete set in $W_{mp}^{os}$. On the contrary, for a non-zero transverse
wave-number $(\beta\neq0)$, the eigenmodes $\tilde{\mathbf{u}}^{+,l}$
of $E_{mp}^{+}$ have both an OS component $\mathbf{u}_{os}^{+,l}$
and a SQ component $\mathbf{u}_{sq}^{+,l}$, and thus are not orthogonal
to $E_{mp}^{-}$ (because $E_{mp}^{-}$ is a complete set in $W_{mp}^{sq}$).
This non-orthogonality of the velocity eigenmodes allows for the possibility
of initial transient growth, that in many cases overshadows the asymptotic
behaviour predicted by the eigenmodes (Butler and Farrell\cite{Butler1992}).
For transverse wave-number $(\alpha=0)$, the non-orthogonality (defined
more precisely in Section \ref{sub:non-orthogonal}) of $E_{mp}^{+}$
with $E_{mp}^{-}$ is important, because in that case the transfer
of energy to the SQ velocity (proportional to $\beta/k^{2}$) is maximum.

\subsection{Transient growth\label{sub:Transient-growth}}

As explained in Section \ref{sub:Temporal-stability-analysis}, the
OS projection $\mathbf{u}_{os}$ of a perturbation $\mathbf{u}^{mp}$
can initiate transient growth. In the following, we will consider
an initial OS perturbation $\mathbf{u}(t=0)=\mathbf{u}_{os}$, in
a linearly stable problem where all the eigenvectors in $E_{mp}$
are damped. From the previous analysis, the perturbation $\mathbf{u}$
can be decomposed into the sum of two contributions: $\mathbf{u}^{+}$
that is a sum of eigenvectors $\tilde{\mathbf{u}}^{+,l}\in E_{mp}^{+}$
and $\mathbf{u}^{-}$ that is a sum of eigenvectors $\mathbf{\tilde{u}}^{-,l}\in E_{mp}^{-}$.

\begin{figure}
\begin{centering}
\subfloat[$t=0$]{\begin{centering}
\resizebox{0.4\textwidth}{!}{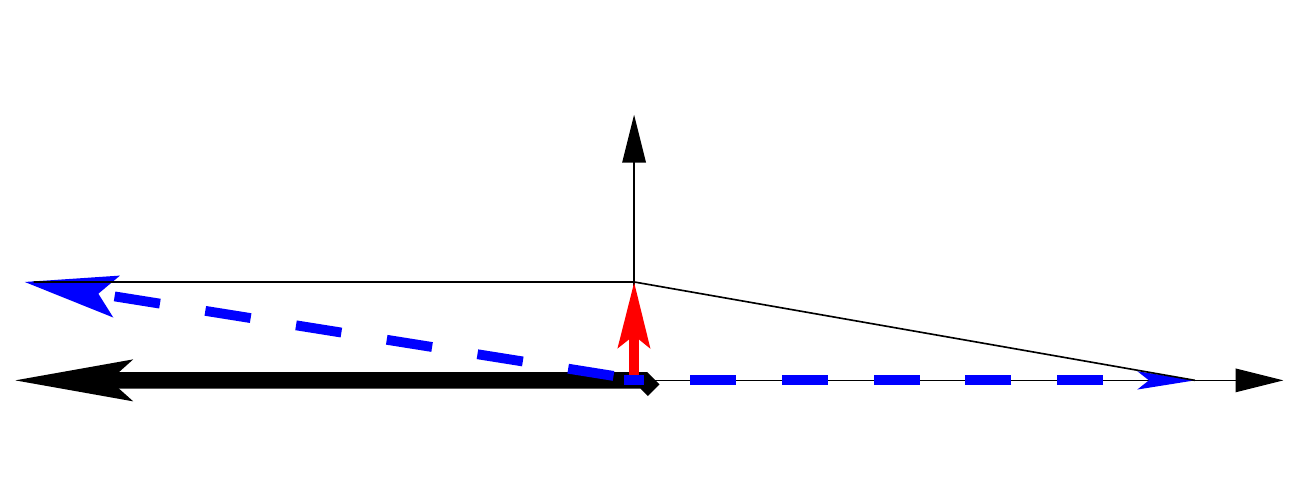}
\par\end{centering}

}\hfill{}\subfloat[$t=t_{0}$]{\begin{centering}
\resizebox{0.4\textwidth}{!}{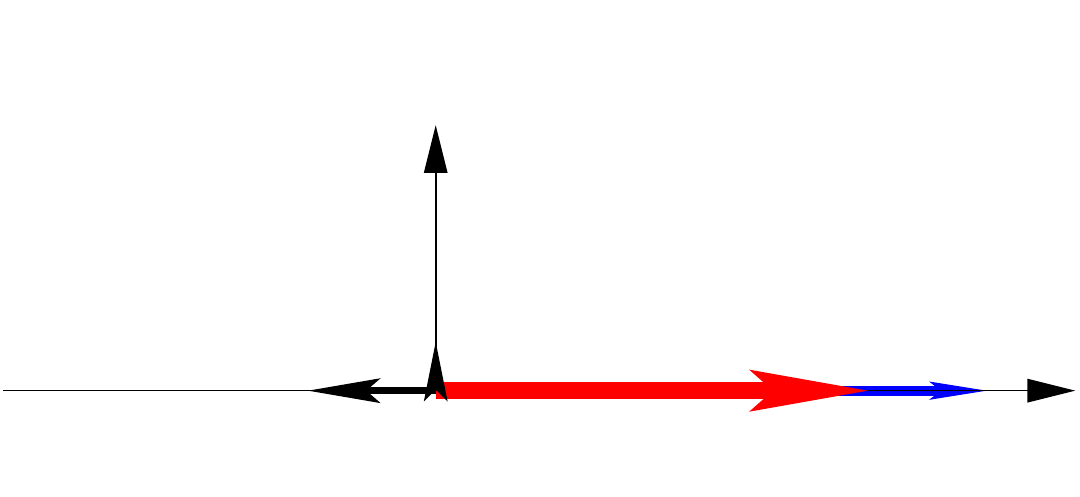}
\par\end{centering}

}
\par\end{centering}

\caption{Sketch illustrating transient growth of an OS perturbation $\bm{u}$.
$\mathbf{\tilde{u}}^{+}$ and $\mathbf{\tilde{u}}^{-}$ are the eigenvectors,\foreignlanguage{french}{
$\mathbf{u}_{os}$} and\foreignlanguage{french}{ $\mathbf{u}_{sq}$}
the projections on $W_{mp}^{os}$ and $W_{mp}^{sq}$: \protect \\
(a) initial perturbation $\bm{u}(0)=\mathbf{u}_{os}^{+}$; (b) response,
at $t=t_{0}$,$\bm{u}(t_{0})=\mathbf{u}_{sq}^{-}$.\label{fig2}}
\end{figure}

To simplify the analysis, let us consider the case where $\mathbf{u}^{+}$
is equal to an eigenvector $\tilde{\mathbf{u}}^{+,l}$:

\begin{equation}
\mathbf{u}=\mathbf{u}^{+}+\mathbf{u}^{-}=\mathbf{u}_{os}^{+,l}+\mathbf{u}_{sq}^{+,l}+\mathbf{u}_{sq}^{-}\mbox{ with at\,}t=0\,\,\,\mathbf{u}_{sq}^{+,l}+\mathbf{u}_{sq}^{-}=0\label{eq11}
\end{equation}
At $t=0$, the SQ projection $\mathbf{u}_{sq}^{+,l}$ of $\mathbf{u}^{+}$
can be decomposed into the set $E_{mp}^{-}$:

\begin{equation}
\mathbf{u}_{sq}^{+,l}=\sum_{n}a_{n}\mathbf{u}_{sq}^{-,n}=-\mathbf{u}_{sq}^{-}\label{eq12}
\end{equation}
The evolution of the perturbation with time is then given by:

\begin{equation}
\mathbf{u}(t)=\mathbf{u}_{os}^{+,l}e^{-\imath\lambda_{l}t}+\sum_{n}a_{n}\mathbf{u}_{sq}^{-,n}\left(e^{-\imath\lambda_{l}t}-e^{-\imath\lambda_{n}t}\right)\label{eq13}
\end{equation}
for non resonant modes, i.e. $\lambda_{l}\neq\lambda_{n}$. Transient
growth can be expected from this relation, if at $t=0$ we have $\left\Vert \mathbf{u}_{os}^{+,l}\right\Vert \ll\left\Vert \mathbf{u}_{sq}^{+,l}\right\Vert $
and if the decomposition \eqref{eq12} of $\mathbf{u}_{sq}^{+,l}$
has large contributions from eigenvectors $\mathbf{u}_{sq}^{-,n}$
with eigenvalues $\lambda_{l}$ different from $\lambda_{l}$, as
illustrated on Figure \ref{fig2}. At $t=0$, the perturbation is
a small OS velocity $\mathbf{u}_{os}^{+}(t=0)$, but consists of a
sum of two large non-orthogonal eigenvectors $\tilde{\mathbf{u}}^{+}$
and $\tilde{\mathbf{u}}^{-}$, having large but opposite SQ velocity.
If the imaginary part of $\lambda^{+}$ is smaller than those of $\lambda^{-}$,
that is the decay rate of $\tilde{\mathbf{u}}^{+}$ is larger than
$\tilde{\mathbf{u}}^{-}$, then $\tilde{\mathbf{u}}^{+}$ decreases
more rapidly than $\tilde{\mathbf{u}}^{-}$. Over time, the perturbation
becomes essentially a SQ velocity $\mathbf{u}_{sq}^{-}(t=t_{0})$,
that can be much larger than the initial SQ velocity $\mathbf{u}_{os}^{+}(t=0)$
for short time $t_{0}>0$. This is however a transient growth, because
in the large time limit $t_{0}\rightarrow\infty$, the perturbation
$\mathbf{u}$ will decrease to zero. A characterisation of this transient
growth is the amplification rate that can be written for an OS perturbation
as:

\begin{equation}
G_{mp}(t_{0})=\frac{\left\Vert \mathbf{u}^{mp}(t_{0})\right\Vert ^{2}}{\left\Vert \mathbf{u}^{mp}(0)\right\Vert ^{2}}\approx\frac{\left\Vert \mathbf{u}_{sq}^{+}(t_{0})+\mathbf{u}_{sq}^{-}(t_{0})\right\Vert ^{2}}{\left\Vert \mathbf{u}_{os}^{+}(0)\right\Vert ^{2}}\label{eq11a}
\end{equation}

\subsection{Optimal mode\label{sub:Optimal-mode}}

A classical tool to analyse transient growth is the determination
of optimal perturbations, that are initial conditions, that will reach
the maximum possible amplification at a given time $t_{0}$. The optimal
mode is the initial perturbation $\mathbf{u}^{mp}(t=0)$ with unity
$L_{2}$-norm having the largest $L_{2}$-norm at time $t_{0}$. By
using the orthogonal decomposition \eqref{eq1}, the maximum possible
amplification $G_{max}(t_{0})$ is:

\[
G_{max}(t_{0})=\max_{\mathbf{u}^{mp}(0)\neq0}\frac{\left\Vert \mathbf{u}^{mp}(t_{0})\right\Vert ^{2}}{\left\Vert \mathbf{u}^{mp}(0)\right\Vert ^{2}}=\max_{\mathbf{u}^{mp}(0)\neq0}\frac{\left\Vert \mathbf{u}_{os}^{mp}(t_{0})\right\Vert ^{2}+\left\Vert \mathbf{u}_{sq}^{mp}(t_{0})\right\Vert ^{2}}{\left\Vert \mathbf{u}_{os}^{mp}(0)\right\Vert ^{2}+\left\Vert \mathbf{u}_{sq}^{mp}(0)\right\Vert ^{2}}
\]

As the transient growth is related to the transfer term in the Squire
equation, and therefore to the growth of the streamwise velocity component,
we expect that the optimal perturbation has initially nearly zero
streamwise velocity component, i.e. $\mathbf{u}_{sq}^{mp}(0)\approx0$,
and thus consists mainly of the Orr-Sommerfeld velocity $\mathbf{u}_{os}^{mp}$.
As this velocity $\mathbf{u}_{os}^{mp}$ induces large streamwise
velocity component and decreases with time, the perturbation $\mathbf{u}^{mp}(t_{0})$
becomes overtime a Squire velocity\foreignlanguage{french}{ $\mathbf{u}^{mp}(t_{0})\approx\mathbf{u}_{sq}^{mp}(t_{0})$}.
Thus the maximum possible amplification $G(t_{0})$ should verify:

\[
G_{max}(t_{0})\approx\max_{\mathbf{u}_{os}^{mp}(0)\neq0}\frac{\left\Vert \mathbf{u}_{sq}^{mp}(t_{0})\right\Vert ^{2}}{\left\Vert \mathbf{u}_{os}^{mp}(0)\right\Vert ^{2}}
\]

\subsection{Resonance}

Several authors (Hultgren et al\cite{Hultgren1981}, Schmid and Henningson\cite{Schmid2001}
and Zaki and Durbin\cite{Zaki2005}) have considered the possibility
of degenerate eigenvalues between Orr Sommerfeld and Squire eigenmodes
to explain fast transient growth through a resonance. Using the decomposition\eqref{eq1},
we will demonstrate that a modal degeneracy between two eigenvectors
$\tilde{\mathbf{u}}^{+,l}$ and $\tilde{\mathbf{u}}^{-,l}$ is impossible
because $\mathbf{\tilde{u}}^{+,l}$ has a non-zero OS projection $\mathbf{u}_{os}^{+,l}$
orthogonal to $\tilde{\mathbf{u}}^{-,l}$.

Indeed, suppose that a modal degeneracy exists, then 2 eigenvectors
$\tilde{\mathbf{u}}^{+,l}=\mathbf{u}_{os}^{+,l}+\mathbf{u}_{sq}^{+,l}$
in $E_{mp}^{+}$ and $\tilde{\mathbf{u}}^{-,l}=\mathbf{u}_{sq}^{-,l}$
in $E_{mp}^{-}$ share the same eigenvalue$\lambda$. Then, because
of the coupling operator $\mathcal{C}_{sq}$ in the eigenvalue problem
\eqref{eq9}, the vector $\tilde{\mathbf{u}}^{+,l}+\theta\mathbf{\tilde{u}}^{-,l}=\mathbf{u}_{os}^{+,l}+\mathbf{u}_{sq}^{+,l}+\theta\mathbf{u}_{sq}^{-,l}$
is also an eigenvector in $E_{mp}^{+}$ associated with the same eigenvalue$\lambda$.
Thus an infinite number of eigenvectors $\mathbf{u}_{os}^{+,l}+\mathbf{u}_{sq}^{+,l}+\theta\mathbf{u}_{sq}^{-,l}$
($\theta\in\mathbb{R}$) share the same eigenvalue $\lambda$. As
they are not countable, they should be dependant as their Squire projections.
Thus the two vectors $\mathbf{u}_{sq}^{+,l}$ and $\mathbf{u}_{sq}^{-,l}$
should be linearly dependant, which implies that $\mathbf{u}_{sq}^{+,l}$
is an eigenvector in $E_{mp}^{-}$ associated with the eigenvalue
$\lambda$. This implies that the transfer term in \eqref{eq9} must
be zero, i.e. $<\frac{\bm{\imath}\beta}{k^{2}}\left(v_{os}^{\text{+,l}}d_{y}U^{b}\right),\eta>=0\,\,\forall\eta$
and thus the OS projection $\mathbf{u}_{os}^{+,l}=0$ (because $\beta\neq0$),
which is inconsistent with the initial assumption.

Thus in bounded domains, transient growth of disturbances cannot be
attributed to exact resonance of Squire modes with Orr-Sommerfeld
modes. However, the eigenvectors $\tilde{\mathbf{u}}^{+,l}$ and $\tilde{\mathbf{u}}^{-,l}$
can have close eigenvalues $\lambda_{l}^{+}$ and $\lambda_{l}^{-}$
and thus the possibility of near resonance still exists.

\section{Numerical solution for wall-bounded flow\label{sec:Numerical-solution}}

The wall-bounded flow studied here consists of thin boundary layers
developing between two parallel walls at large Reynolds number. The
channel height is such that the boundary layer thickness is small
compared to the wall distance, so that there is no interaction between
the two boundary layers. Such a configuration was used by Mack\cite{MackJFM76}
to study the eigenvalue spectrum of the Blasius boundary layer. This
configuration corresponds also to the experimental set-up used in
wind tunnels to study boundary layer. However, in numerical simulations,
this approach is seldom used and alternative approaches are usually
preferred, such as mapping transformation (Fisher\cite{fisherNumMath09})
or direct numerical integration (Jacobs and Durbin\cite{Jacobs1998}),
but imposition of boundary conditions at infinity may remain problematic.

At large Reynolds number, the considered base flow is a nearly parallel
mean flow $\mathbf{U}^{b}\approx U^{b}(y)\,\mathbf{x}$, corresponding
to a Blasius profile in each half of the domain. The Reynolds number
in the channel, $Re_{h}=U_{0}h/\nu$, is equal to $20\,000$. The
analysed section is located at $x=2h$ from the entrance, that corresponds
to a boundary layers thickness $\delta/h=0.05$ and a Reynolds number
(based on the displacement thickness$\delta_{1}$) $Re_{\delta_{1}}=344$.
This Reynolds number is lower than the critical Reynolds number $Re_{\delta_{1}}=520$,
such that, for the considered case, all the eigenmodes are decaying
with time. Dimensionless quantities with respect to the displacement
thickness $\delta_{1}$ are denoted by an asterisk. To solve the variational
eigenvalue problem \eqref{eq9}, we use a spectral Galerkin method
with Chebyshev approximation described in Buffat et al\cite{Buffat2009}.
The corresponding NadiaSpectral computer code has been validated in
Buffat et al\cite{Buffat2009} by comparison to linear stability analysis
of plane Poiseuille flow. Using Chebyshev polynomials of order $N_{y}=192$
insure a relative error of at least $10^{-12}$ for the first$N_{y}$
eigenvalues of the system \eqref{eq9} (of size $2N_{y}$).

\subsection{Optimal mode}

Using the variational formulation of Butler and Farrell\cite{Butler1992}
with the eigenvalue problem \eqref{eq7}, the optimal perturbation
has been calculated at $Re_{\delta_{1}}=344$ . The transient time
$t_{0}$ is equal to the time $t_{max}$, at which the transient growth
is maximum for a Blasius boundary layer with $\alpha=0$, i.e. $t_{max}=0.8\, Re_{\delta_{1}}\delta_{1}/U_{0}$
(Butler and Farrell\cite{Butler1992}). The maximum value of the amplification
$G_{max}(t_{max})$ is obtained for a spanwise wavenumber $\alpha^{*}=0$,
$\beta^{*}\approx0.67$ and $G_{max}(t_{max})$ reaches $186$. These
values are closed to the values found by Butler and Farrell\cite{Butler1992})
for the boundary layer ($\beta^{*}=0.65$ and $G_{max}(t_{max})=177$).
We observe also that the transient growth rate is very large, i.e.
greater than $100$, for a large range of disturbances having spanwise
wavenumber $\beta$ between $1/\delta$ and $3/\delta$ and streamwise
wavenumber at least ten times smaller $\alpha\leq0.1/\delta$.

\begin{figure}
\begin{centering}
\includegraphics[width=0.45\textwidth]{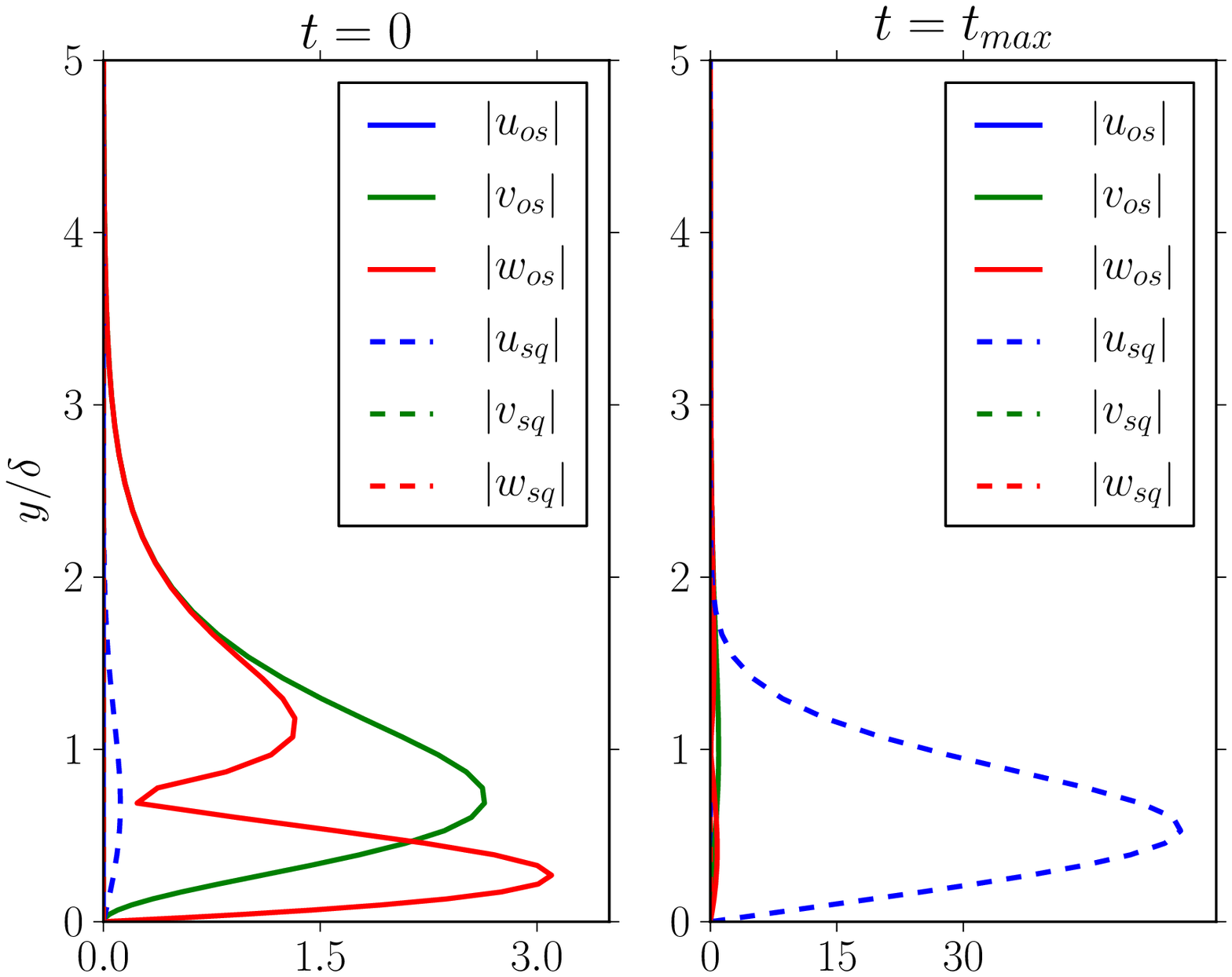}\includegraphics[width=0.45\textwidth]{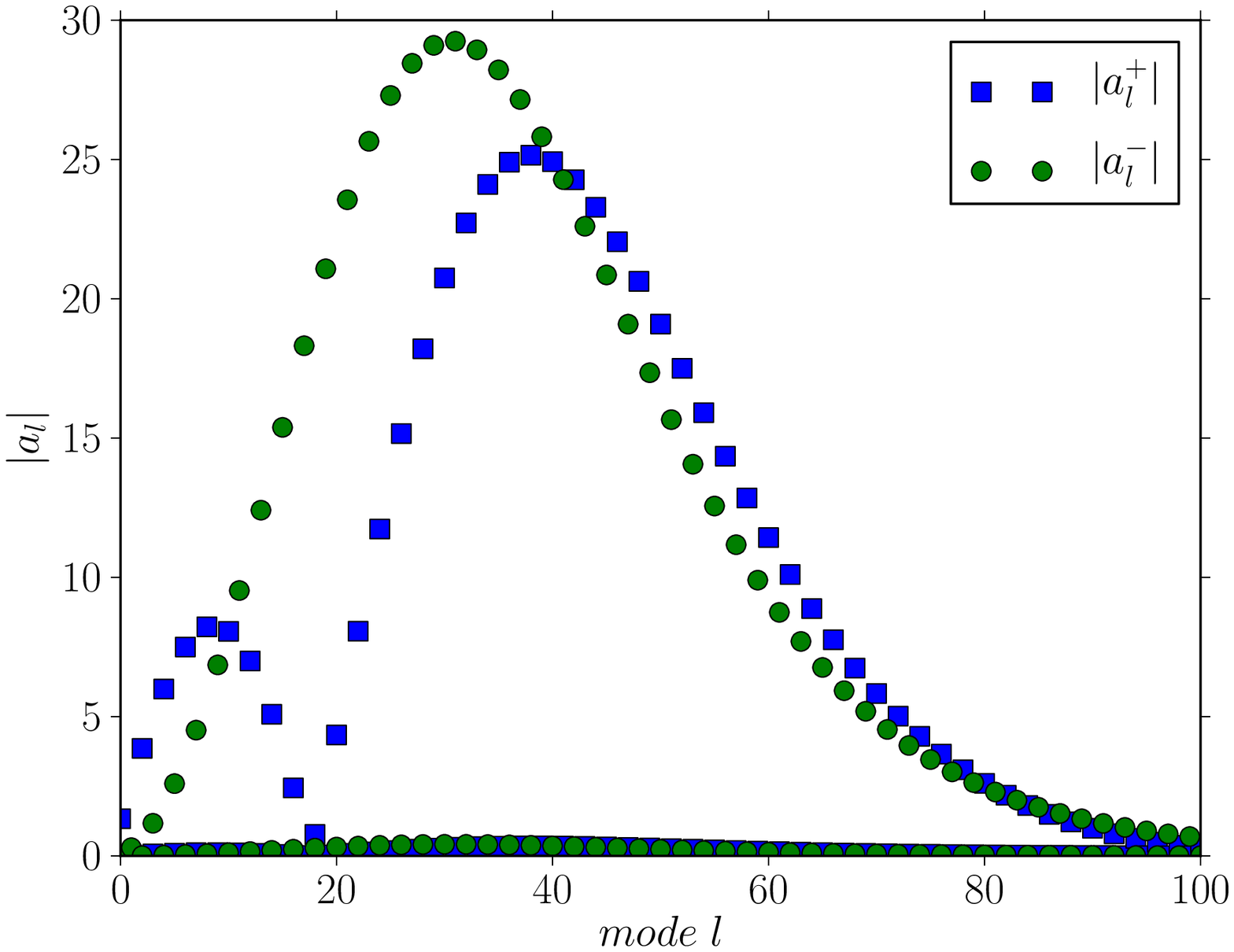}
\par\end{centering}

\caption{Optimal perturbation at $Re_{\delta}=344$ for a spanwise wave $\alpha^{*}=0$,$\beta^{*}=0.67$:
a) (left) norm of the optimal perturbation with an OS-SQ decomposition;
b) (right) modulus of the expansion coefficients $a_{l}^{+}$ and
$a_{l}^{-}$ versus the mode number $l$ sorted by decreasing eigenvalue
imaginary part.\label{fig3}}
\end{figure}

Figure \ref{fig3}a shows the calculated optimal mode and its orthogonal
decomposition at $t=0$ and $t=t_{max}$. As seen on this figure,
the initial optimal disturbance is an OS velocity characterised by
nearly spanwise vortices inside the boundary layer. Then, as expected,
the perturbation transforms itself over time into Squire velocity,
and at $t=t_{max}$ this perturbation is mainly a SQ velocity, that
presents a large peak in the streamwise direction. This profile corresponds
to the classical shape of streaks inside the boundary layer. Figure
\ref{fig3}b shows the repartition of the modulus of the expansion
coefficients $a_{l}$ in the eigenvectors basis $E^{+}$ and $E^{-}$
for this optimal perturbation:

\[
\mathbf{u}_{opt}(t)=\sum_{l}a_{l}^{+}e^{-\imath\lambda_{l}t}(\mathbf{u}_{os}^{+,l}+\mathbf{u}_{sq}^{+,l})+\sum_{n}a_{n}^{-}e^{-\imath\lambda_{n}t}\mathbf{u}_{sq}^{-,n}
\]
As seen in this figure, the optimal mode is a wide combination of
eigenmodes $\tilde{\mathbf{u}}^{+,l}$ and $\tilde{\mathbf{u}}^{-,l}$.
In the following, the mode number $l$ is sorted by decreasing eigenvalue
imaginary part, such that a low mode number $l$ corresponds to an
eigenvalue $\tilde{\mathbf{u}}^{l}$ with a low decay rate. We notice
also the clear separation between the dominant modes associated with
the coefficients $a_{l}^{+}$ and $a_{l}^{-}$, indicating that they
are associated with well-separated eigenvalues. The modulus of the
coefficients is large ($\gg1$), indicating a strong non-orthogonality
of the eigenvectors (for orthogonal eigenvectors we should have $|a_{l}|<1$).
As the coefficients $a^{-}$ are larger than $a^{+}$ for small mode
numbers $l$, they are associated with eigenmodes with smaller decay
rate and the optimal mode will contain at $t=t_{max}$ essentially
$\mathbf{u}_{sq}^{-,l}$ eigenvectors:

\[
\mathbf{u}_{opt}(t_{max})\approx\sum_{n}a_{n}^{-}e^{-\imath\lambda_{n}t_{max}}\mathbf{u}_{sq}^{-,n}
\]

\subsection{Transient growth\label{sub:non-orthogonal}}

To characterize the link between transient growth and non-orthogonality,
we define for each eigenvector $\mathbf{u}^{+,l}$ a coefficient $\Xi_{mp}^{l}$
defined as the relative norm of its OS projection:

\[
\Xi_{mp}^{l}=\frac{\left\Vert \mathbf{u}_{os}^{+,l}\right\Vert _{L^{2}}}{\left\Vert \mathbf{u}^{+,l}\right\Vert _{L^{2}}}
\]
This coefficient is the absolute value of the cosine of the angle
between the eigenvector and the OS velocity space and thus characterises
the orthogonality of the eigenvector with the set $E_{mp}^{-}$. If
for all eigenvectors $\Xi_{mp}^{l}\approx1$, then the set $E_{mp}^{+}$
is orthogonal to $E_{mp}^{-}$ and transient growth does not occur.
On the other hand, transient growth is possible, with an initial perturbation
equal to the OS projection of $\mathbf{u}^{+,l}$, if $\Xi_{mp}^{l}\approx0$
for some eigenvectors $\mathbf{u}^{+,l}$ with low decay rate (i.e.
associated with low mode number $l$).

\begin{figure}
\begin{centering}
\subfloat[$\Xi_{mp}^{l}$ for $\mathbf{u}^{+,l}$ eigenvalues]{\begin{centering}
\includegraphics[width=0.45\textwidth]{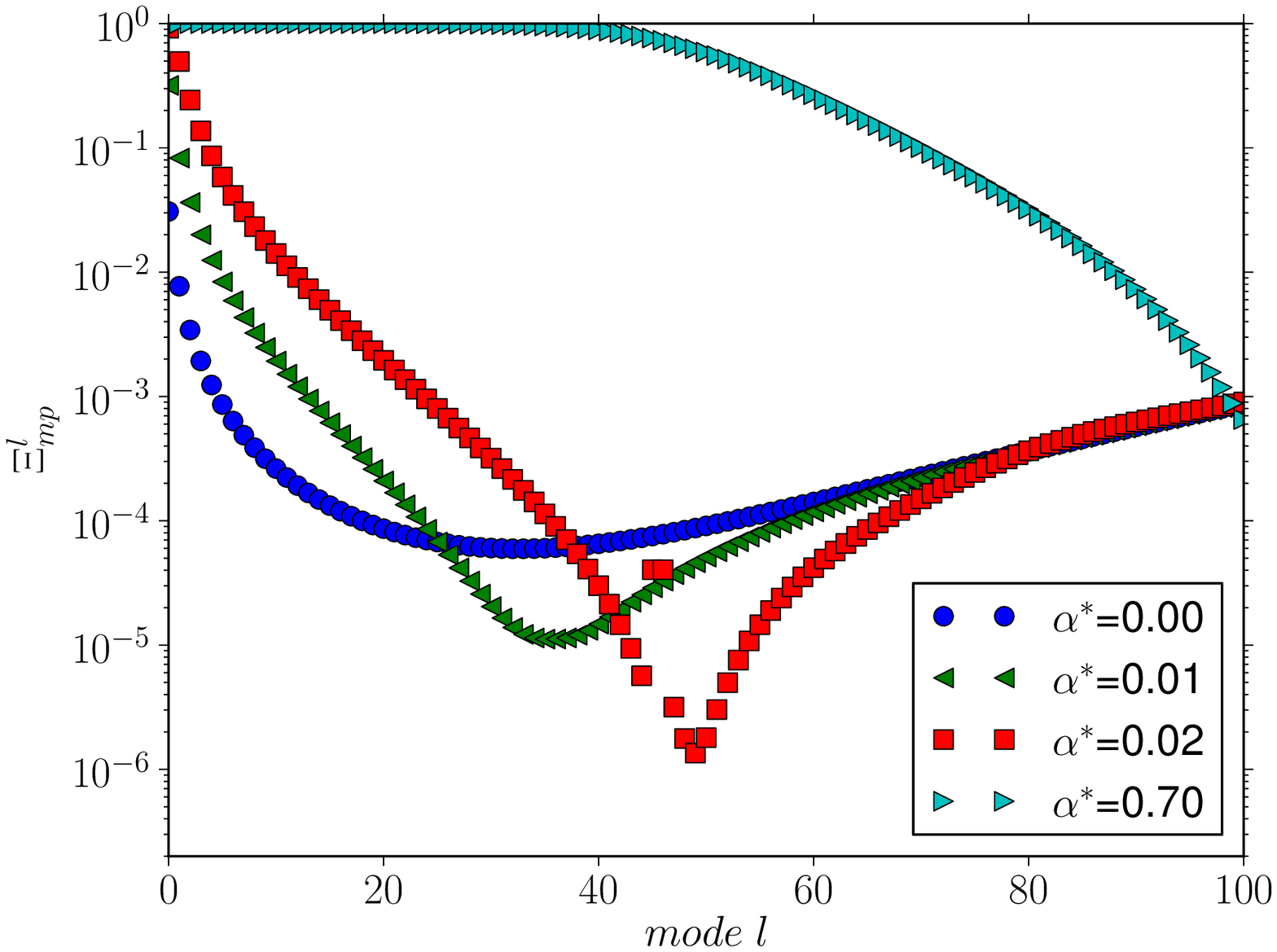}
\par\end{centering}

}\subfloat[transient growth $G_{mp}^{l}$ of $\mathbf{u}_{os}^{+}$]{\begin{centering}
\includegraphics[width=0.45\textwidth]{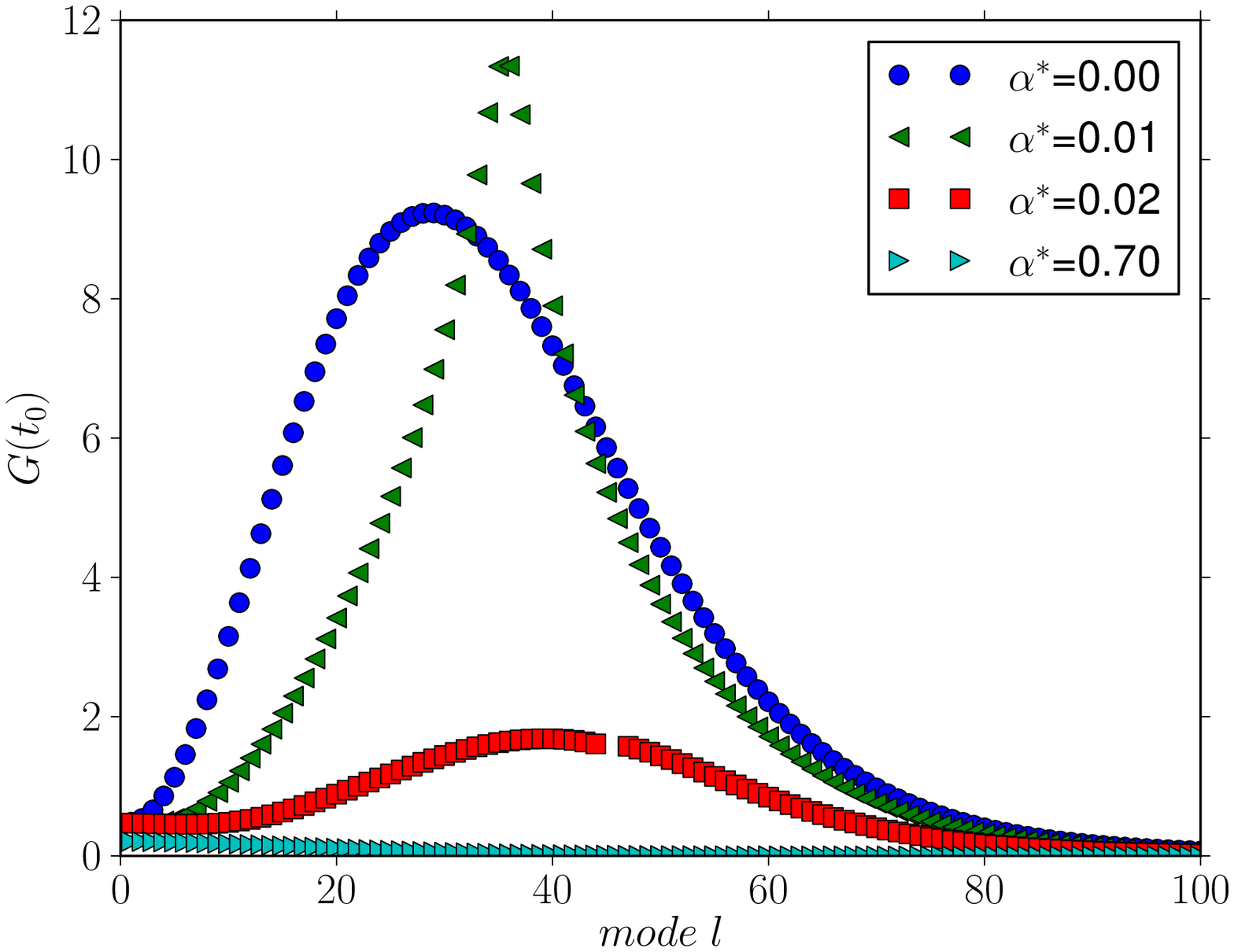}
\par\end{centering}

}
\par\end{centering}

\caption{a) Orthogonality coefficient $\Xi_{mp}^{l}$ and b) amplification
rate $G_{mp}(t_{max})$ at $t_{0}=t_{max}$ for different wavenumber
$\alpha^{*}$versus the mode number $l$ sorted by decreasing eigenvalue
imaginary part ($Re_{\delta}=344$, $\beta^{*}=0.67$)\label{fig4}}
\end{figure}

The value of this coefficient $\Xi_{mp}^{l}$ and the corresponding
transient growth $G_{mp}^{l}$ are plotted in Figure \ref{fig4} as
a function of the mode number $l$ and for various streamwise wavenumber
$\alpha^{*}$. The spanwise wavenumber is $\beta^{*}=0.67$, but similar
plots are obtained for $\beta$ between $1/\delta$ and $3/\delta$.
As seen on this figure, for zero or low values of $\alpha^{*}$, the
coefficient $\Xi_{mp}^{l}$ is very small for a large number of modes
with low mode numbers $l$, that are associated with large transient
growth. On the contrary for larger value of $\alpha^{*}$, the coefficient
$\Xi_{mp}^{l}$ remains equal to one for low mode numbers and no transient
growth is observed.

For zero streamwise wavenumber $\alpha^{*}=0$, analytical solutions
can be obtained for $\mathbf{u}_{sq}^{-,l}$ and $\mathbf{u}_{os}^{+,l}$
as in Drazin and Reid\cite{drazin04}. Outside the boundary layer,
they are sinusoidal functions in the wall normal direction $y$ with
a wavenumber $\mu$, independent of the Reynolds number, and solution
of transcendental equations: $\mu_{l}^{+}\cot\mu_{l}^{+}=\beta\coth\beta$
and $\mu_{l}^{-}=l\pi$ for odd modes, $\mu_{l}^{+}\tan\mu_{l}^{+}=-\beta\tanh\beta$
and $\mu_{l}^{-}=(2l+1)\pi/2$ for even modes. It can be found that
only the projection $\mathbf{u}_{sq}^{+,l}$ depends on the gradient
of the mean flow and scales with the Reynolds number. Thus for $\alpha=0$,
the coefficient $\Xi_{mp}^{l}$ scales at high Reynolds number as
the inverse of the Reynolds number, indicating as expected an increase
of transient growth with the Reynolds number. For low value of the
streamwise wavenumber $\alpha$, the coefficient $\Xi_{mp}^{l}$ is
very small for many eigenmodes with low mode number $l$, whereas
for larger value of $\alpha$ the coefficient $\Xi_{mp}^{l}$ remains
equal to one for small $l$ as seen in Figure \ref{fig4}. By looking
at the shape of the eigenvectors, we conclude that the eigenmodes
with very small $\Xi_{mp}^{l}$ and low mode number $l$ correspond
to eigenmodes $\mathbf{\tilde{u}}^{+,l}$, having an half-wavelength
$\pi/\mu$ in the normal direction $\mathbf{e}_{y}$ of the order
of the boundary layer thickness $\delta$.

\section{Discussion\label{sec:Discussion}}

Using the orthogonal decomposition \eqref{eq1} of the velocity perturbation,
we have a clear demonstration of the link between transient growth
and the non-orthogonality of the eigenvectors of the Orr-Sommerfeld
Squire equations. Large transient growth results from the non-orthogonality
of the two sets of velocity eigenmodes $E_{mp}^{+}$ and $E_{mp}^{-}$
and the ability of an OS velocity to transfer energy to a SQ velocity.
To generate large transient growth, a general perturbation (as a free-stream
turbulence) must induce large transfer of kinetic energy to the SQ
velocity. This perturbation must thus contain OS velocity $\mathbf{u}_{os}$
associated with non-orthogonal eigenmodes $\mathbf{u}^{+,l}$ with
low decay rate (i.e. such that $\Xi_{mp}^{l}\approx0$ for small $l$).
They are essentially streamwise vortex $(\alpha\ll\beta)$ with a
spanwise half-wavelength and a normal half-wavelength of the order
of the shear distance (the boundary layer thickness). Over time this
OS velocity perturbation can develop large SQ velocity perturbation.
This transient growth problem is similar to the initial value problem
considered in Zaki and Durbin\cite{Zaki2005} to study boundary layer
transition due to free-stream turbulence. From the Squire equation
they consider the initial value problem for the case of Squire modes,
generated by a single Orr\textendash{}Sommerfeld mode forcing. They
define a coupling coefficient to characterize the ability of Orr\textendash{}Sommerfeld
mode to generate large Squire response. The low frequency Orr-Sommerfeld
mode with large coupling coefficient are called ``penetrating modes'',
and they correspond to OS velocity $\mathbf{u}_{os}$ associated with
non-orthogonal eigenmodes $\mathbf{u}^{+,l}$ with low decay rate.
Zaki and Durbin\cite{Zaki2005} attribute the large growth of disturbances
generated by theses penetrating modes to exact resonance between Squire
and Orr-Sommerfeld modes. Existence of resonance for a boundary layer
in a semi-infinite domain is invoked by Zaki and Durbin\cite{Zaki2005},
arguing that since the dispersion relation for the temporal continuous
spectrum modes being identical for the Orr-Sommerfeld and Squire modes,
they can have identical eigenvalues. As demonstrated in section \ref{sec:Wall-bounded-shear},
this is not true in a bounded domain where exact resonance is impossible
between the two sets of velocity eigenmodes $E^{+}$ and $E^{-}$.
However, eigenvectors in the two sets can have close eigenvalues and
thus the possibility of near resonance still exists. Thus, in bounded
domains and presumably in infinite domains also, large transient growth
are mainly the consequence of the non-orthogonality between the two
sets $E^{+}$ and $E^{-}$ of velocity eigenmodes.

Destabilizing perturbations $\mathbf{u}_{os}$, that lead to large
transient growth, are OS projection $\mathbf{u}_{os}^{+}$ of eigenvectors
$\mathbf{u}^{+,l}$ with $\alpha\ll\beta$ and $\pi/\beta\approx\delta$
(i.e with $\Xi_{mp}^{l}\approx0$ for small $l$). Their transient
growth can trigger the boundary layer transition induced by free-stream
turbulence. Indeed as the two sets $E^{+}$ and $E^{-}$ form a complete
set, any free-stream turbulence can be expanded using these two sets.
The modes in the free-stream turbulence that trigger the first instability
are the destabilizing perturbations $\mathbf{u}_{os}^{+}$, that create
streaks. To initiate the destabilisation of these streaks, higher
frequency perturbations in the free-stream induce inflectional instability
that leads to turbulent transition (Zaki and Durbin\cite{Zaki2005},
Schlatter et al\cite{Schlatter2008}).

The number of destabilizing perturbations $\mathbf{u}_{os}^{+}$ is
large and a particular combination can be formed to optimize the transient
growth. This is the optimal mode $\mathbf{u}_{opt}$, that combines
the perturbations $\mathbf{u}_{os}^{+}$ such that $\mathbf{u}_{opt}$
is nearly zero outside the shear region because in that region the
transfer term is zero. This optimal mode can model perturbations inside
the boundary layer, like spanwise periodic array of small cylindrical
roughness elements fixed on the wall (Fransson et al\cite{Fransson2004}).

A remarkable fact is that the shape of the transient response is nearly
identical for an initial condition equal to the optimal mode $\mathbf{u}_{opt}$
and for initial conditions equal to the OS projections $\mathbf{u}_{os}^{+}$
of a large number of eigenvectors $\mathbf{u}^{+,l}$. Only the amplitude
of the transient response depends on the particular initial condition.
Some considerations supporting this expectation are given in Appendix
A.

\section*{Appendix A. Shape of the transient response}

For a wide range of initial conditions, the transient response corresponds
to streaks characterised by a large peak of streamwise velocity component
inside the boundary layer. The generic shape of this peak, characteristic
of the transient response, can be explained by looking at the equation
for the SQ streamwise velocity $u_{sq}$ obtained from the weak formulation
\eqref{eq2b} and the decomposition \eqref{eq1}:

\begin{equation}
\left(\partial_{t}+\bm{\imath}\alpha U^{b}\right)u_{sq}+\frac{\beta^{2}}{k^{2}}d_{y}U^{b}v_{os}\,-Re^{-1}\mathcal{D}_{mp}^{2}u_{sq}=0\label{eq15}
\end{equation}
As large transient growth corresponds mainly to the SQ velocity $\mathbf{u}_{sq}^{-,l}$
with eigenvalues near the eigenvalue $\lambda$ of the perturbation
$\mathbf{u}_{os}^{+,k}$, we are looking for solutions of \eqref{eq15}
of the form:

\[
u_{sq}=F(y)\, u_{sq}^{-,l}t\, e^{-\imath\lambda t}
\]
Taking into account that $u_{sq}^{-,l}$ is an eigenvector of \eqref{eq15}
with $v_{os}=0$, $F(y)$ is a solution of the following equation:

\[
u_{sq}^{-,l}F+\frac{\beta^{2}}{k^{2}}d_{y}U^{b}v_{os}^{+,k}-Re^{-1}t\, u_{sq}^{-,k}d_{y}^{2}F=0
\]
At short time, by neglecting the viscous term proportional to $Re^{-1}$,
we obtain an approximate form for $F$ and thus for the transient
response $u_{sq}$:

\[
u_{sq}\approx-\frac{\beta^{2}}{k^{2}}v_{os}^{+,k}d_{y}U^{b}\, t\, e^{-\imath\lambda t}
\]
As the mean shear $\partial_{y}U^{b}$ is zero outside the boundary
layer, the streamwise transient response $u_{sq}$ is zero outside
the boundary layer and depends on the normal velocity $v_{os}^{+}$
inside the boundary layer. Transient growth is obtained with perturbations
$\mathbf{u}_{os}^{+,k}$ associated with small decay rates and normal
wavelengths of the order of a few boundary layer thickness $\delta$.
In that case $v_{os}^{+,k}\approx Cy/\delta$ in the boundary layer,
and an approximation for the transient response reads:

\begin{equation}
u_{sq}\approx-\frac{\beta^{2}}{k^{2}}C\frac{y}{\delta}d_{y}U^{b}\, t\, e^{-\imath\lambda t}\label{eq17}
\end{equation}
In that case the shape (along $y$) $y\, d_{y}U_{b}$ of the transient
response $u_{sq}$ is independent of the destabilizing perturbation
$v_{os}$, and only its amplitude $\frac{\beta^{2}}{k^{2}}\frac{C}{\delta}$
is a function of the perturbation $v_{os}$ ($C$ depends on the normal
wavelength of $v_{os}$).

Luchini\cite{LuchiniJFM00} had already noted that the shape of the
transient response for the optimal initial perturbation is similar
to the shape for more generic initial perturbations. He pointed out
that this shape looks very much like an analytical expression (due
to Stewartson 1957) simply given by $y\, d_{y}U_{b}$.

\bibliographystyle{apsrev4-1long}
\bibliography{BIBLIO/biblio_moveo_nadia,BIBLIO/biblio_methodespectral}

\end{document}